\documentclass[aps,prl,reprint,twocloumn,superscriptaddress,longbibliography]{revtex4-2}

\usepackage{graphicx} % For including figures
\usepackage{float}
\usepackage{braket}   % For \bra{}, \ket{}, \braket{}
\usepackage{comment}  % For \begin{comment}... \end{comment} blocks
\usepackage{color}    % For \textcolor (NOTE: Remove all editorial \textcolor usage for final submission)
\usepackage[normalem]{ulem} % For \sout{} (NOTE: Resolve all \sout usage for final submission)
\usepackage[colorlinks=true,citecolor=blue,linkcolor=magenta,breaklinks=true]{hyperref} % For hyperlinks
\usepackage[utf8]{inputenc}
\usepackage{amsmath,amssymb,physics,bm,mathtools}
\usepackage{bbm}

\begin{document}
\title{Emergent Nodal Spheres and Weyl Fermions via Spin-Texture Coupled to Thin Film Orbital Dirac Semimetals}
\author{Pritam Chatterjee}
\affiliation{Department of Physics, The University of Tokyo, 7-3-1 Hongo, Bunkyo-ku, Tokyo 113-0033, Japan}
\author{Anirudha Menon}
\email{menon.cond.mat@gmail.com}
\affiliation{Centre for Theoretical and Computational Physics, National Yang Ming Chiao Tung University, Hsinchu City, Taiwan}

\date{\today}

\begin{abstract}
We consider the minimal coupling of a thin film Dirac semimetal Hamiltonian to a generic spin-texture. A simple unitary transformation gauges away the spatial dependence in the exchange term, leading to the generation of effective corrections to the Dirac dispersion. A full function's worth of freedom is obtained as a result. Choosing different pitch vectors, we show that many novel phenomena arise in such systems. For example, a linear pitch vector leads to the generation of a Weyl semimetal --- we observe the anomalous Hall effect and the chiral magnetic effect. The anomalous Hall coefficient requires a non-zero pitch vector whereas the CME is proportional to the exchange coupling. The band structure of the model in the presence of a magnetic field shows a Lifshitz-like transition driven by the exchange coupling. The introduction of a suitable time-dependent pitch vector leads, at the level of the leading-order Floquet effective Hamiltonian, to the emergence of a nodal sphere in momentum space. We further show that, in the full driven problem, a closed quasienergy degeneracy structure persists, continuously connected to this nodal sphere, and constrained by the operator algebra of the Floquet expansion.
\end{abstract}

\maketitle

\noindent{\it Introduction.} 
Topological semimetals~\cite{TopoS1,WSM1,WSM2,WSM3,WSM4,DSM1,DSM2,DSM3,DSM4,NLSM1,NLSM2,NLSM3,NLSM4} have been studied extensively by the condensed matter community since their first predictions over a decade ago. This class of materials exhibits point- or line-like degeneracies arising from band touchings between the conduction and valence bands near the Fermi energy. Prominent examples include Dirac semimetals~\cite{DSM1,DSM2,DSM3,DSM4}, Weyl semimetals~\cite{WSM1,WSM2,WSM3,WSM4}, and nodal-line semimetals~\cite{NLSM1,NLSM2,NLSM3,NLSM4}, realized in both spinless and spinful systems. Here, ``spinless'' refers to electronic systems in which spin--orbit coupling (SOC) can be effectively neglected \cite{spinlessDSM1}, rendering spin a passive degeneracy. While these phases have been widely explored using theoretical and computational approaches, not all of them have been connected to experimentally viable material platforms.

Dirac semimetals (DSMs) are characterized by four linearly dispersing bands meeting at a point-like degeneracy known as the Dirac point. Although the total Chern number associated with a Dirac point vanishes identically, the Berry phase accumulated along symmetry-allowed loops encircling the node is quantized to $\pi$. Dirac points are generally unstable against perturbations and may gap out or split into pairs of Weyl points unless protected by crystalline symmetries~\cite{spinlessDSM1}. Spinful DSMs, in which SOC plays a central role, have been studied extensively, with experimental realizations reported in Cd$_3$As$_2$ and Na$_3$Bi. In contrast, orbital DSMs preserve full SU$(2)$ spin-rotation symmetry and have been discussed primarily from the perspective of representation theory~\cite{spinlessDSM2, DSM1}. One reason this class has received comparatively less attention is the absence of a general microscopic guiding principle for realizing band inversion---an essential ingredient for topological semimetals---in the absence of SOC. Moreover, until recently, few material candidates had been proposed for orbital DSMs. Despite these challenges, such DSMs are promising platforms for hosting a range of exotic physical phenomena, as we demonstrate in this work.

Recently, non-collinear magnetic textures and other unconventional magnetic orders have emerged as versatile platforms for engineering fundamental physical phenomena in condensed matter systems and for advancing spintronics applications. Electrons coupled to non-collinear magnetic textures experience emergent (fictitious) gauge fields, including axial electromagnetic fields~\cite{Yasufumi2020,Bruno2004,Schulz2012,Nagaosa2012,Liu2013,Ozawa2024,Ilan2020}. Spatial or temporal variations in (anti)ferromagnetic order lead to a broad range of electronic responses beyond those found in systems with uniform and static magnetism. Such non-collinear spin configurations can give rise to phenomena including the anomalous Hall effect (AHE) and the chiral magnetic effect (CME)~\cite{Terasawa2024,Yasufumi2020,Ishihara2019,An2019,Redies2020,Ilan2020,Heidari2023,Harada2023}. Furthermore, the coupling between emergent gauge fields from magnetic textures and superconducting order parameters has been shown to generate rich physics, including topological superconductivity hosting Majorana end, edge, and corner modes~\cite{Loss2022,chatterjee2024_I,Chatterjee2024_II,Subhadarshini2024,Chatterjee2023,Brüning2025,Subhadarshini2025}. Non-collinear magnetism thus provides an alternative route for engineering chirality and topology in solids without the need for external gauge fields or laser drives.

There has been some work on the interaction of topological semimetals with spin-textures \cite{Araki2018, bera2025} but there are still gaps in the literature. In this work, we consider the effect of spin-textures on orbital DSMs leading to a host of novel phenomena including the appearance of Weyl points, associated transport phenomena, and Floquet-induced degeneracy structures descended from a nodal sphere in the effective Hamiltonian.

\begin{figure}[h]
\centering
\includegraphics[width=0.5\textwidth]{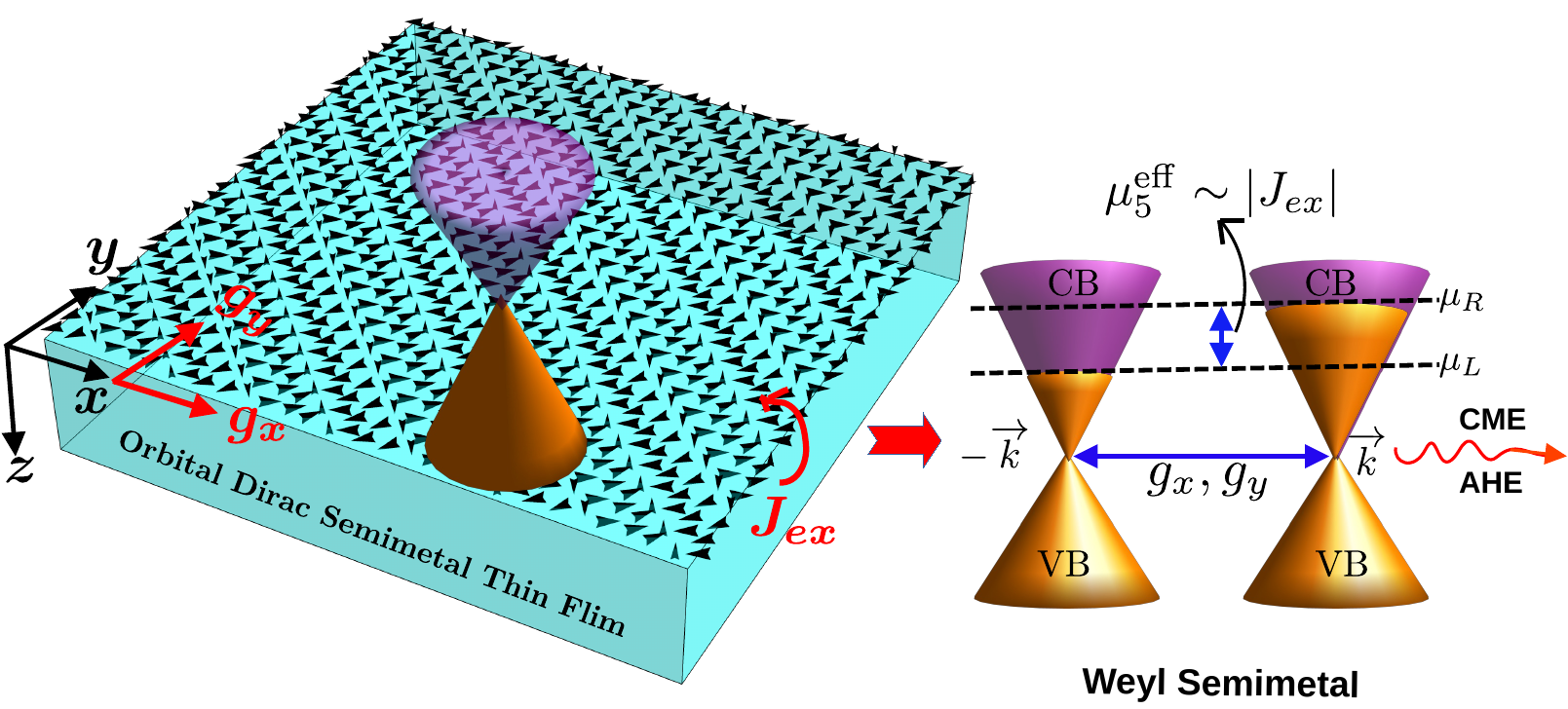}
\caption{Schematic illustrating the signatures of the CME and AHE in an effective Weyl spectrum originating from an orbital Dirac semimetal under spin-texture dynamics. }
\label{Fig1}
\end{figure}

\noindent {\it Model and Unitary Rotation}. The setup is such: we envision a thin film 3d Dirac semimetal sample with a spin-texture on the surface as shown in Fig. \ref{Fig1}. The presence of the thin film allows for the magnetic interaction to permeate the entire sample at roughly constant strength. An orbital Dirac semimetal minimal model Hamiltonian is then considered coupled to a spin-texture exchange term, with the effective Hamiltonian of the form

\begin{align}
    H(\vec{k},\vec{r}) = v_F \sigma_0 (\vec{\tau} \cdot \vec{k}) + J_{ex} \vec{S}(\vec r)\cdot \vec{\sigma} ,
\end{align}

where $\vec{\tau}$ is a valley degree of freedom, $\vec{\sigma}$ is real spin, and $\vec{S}$ represents a classical spin. $v_F$ represents the so called Fermi velocity of the DSM and $J_{ex}$ represents the strength of the spin-texture coupling. Such a model may be derived from microscopic theories of interfaces using the Schreiffer-Wolf class of transformations. We assume the following form for $\vec{S}$:
\begin{align}
\vec{S}(\vec r)
\!\!=\!\!|\vec{S}|
\left[
\sin \theta\cos \phi,\;
\sin \theta\sin \phi,\;
\cos \theta
\right].
\end{align}
in spherical coordinates and set $\theta(r) = \pi/2$ without loss of generality. This leads to a dependence of the $\vec{S}$ vector purely on $\phi(r)$. We choose a form for $\phi$ as $\phi(r) \equiv \phi(x,y)$, such that $\partial_z \phi = 0$. 

A natural unitary transformation presents itself as $U(r) = \exp[-i\frac{\phi(r)}{2}\sigma_z \otimes \tau_0]$ to re-express the Hamiltonian in a convenient form with the exchange term being simplified to a position independent term. Under this transformation, the Dirac piece transforms as

\begin{align}
     U^{\dagger}(r) v_F\tau_ik_i U(r) 
    &= v_F\tau_i U^{\dagger}(r) k_i U(r) \nonumber \\
    &= -iv_F\tau_i U^{\dagger}(r) \partial_i U(r) \nonumber \\
    &= -iv_F\tau_i [ U^{\dagger}(r) U(r) \partial_i \nonumber \\ &+  U^{\dagger}(r) \{\partial_i U(r)\}]\nonumber \\
    & = -iv_F\tau_i [ U^{\dagger}(r) U(r) \partial_i \nonumber \\ &-i  \frac{1}{2} U^{\dagger}(r) U(r) \partial_i \phi(r) \sigma_z] \nonumber \\
    & = v_F\tau_i [ k_i -  \frac{1}{2} \partial_i \phi \sigma_z],
\end{align}

where $U^{\dagger}U=1$ and we've dropped the dependence of $\phi$ on $r$. Naturally, the Dirac term then looks like

\begin{align}
    U^{\dagger}H_{DSM}U &= v_F\sigma_0(\vec{\tau} \cdot \vec{k}) \nonumber \\
    & -\frac{1}{2} \sigma_z [\tau_x \partial_x \phi +\tau_y \partial_y \phi].
\end{align}

Next. we look at the effect of the unitary rotation on the exchange spin-texture piece, which is well known, and it turns out that

\begin{align}
    U^{\dagger}J_{ex} \vec{\sigma} \cdot \vec{S} U = J_{ex} |\vec{S}| \sigma_x.
\end{align}

With this, the full transformed Hamiltonian $H_t \equiv U^{\dagger} H U$ takes the form

\begin{align}
    H_{t} &= v_F\sigma_0(\vec{\tau} \cdot \vec{k}) -\frac{1}{2} \sigma_z [\tau_x \partial_x \phi +\tau_y \partial_y \phi] \nonumber \\
    &J_{ex} |\vec{S}| \sigma_x.
\end{align}

The transformation is remarkable --- we now have a DSM Hamiltonian with some extra terms which may induce novel physics into the system. We are free to choose the form of the pitch vector $\phi$, and indeed different cases have been explored for such vectors in the literature.\\

\noindent{\it Linear Pitch vector Induced WSM}. We may now choose a simple form for the pitch vector as a warm up. We find such a simple choice to be $\phi(r) = g_x x + g_y y$. In this case the derivatives work out trivially and we see that we get a Hamiltonian of the form

\begin{align}
    H_{t} &= v_F\sigma_0(\vec{\tau} \cdot \vec{k}) -\frac{1}{2} \sigma_z [g_x \tau_x +g_y\tau_y] \nonumber \\
    &+ J_{ex} |\vec{S}| \sigma_x.
\end{align}

We set $|\vec{S}| = 1$ for convenience. This Hamiltonian may be diagonalized by the {\it squaring trick} as is well known in the literature. The details can be found in the supplementary material. We find that the eigenvalues take the form

\begin{align}\label{Eq8}
    \epsilon_{\pm,\rho}(\mathbf{k})
    = \rho\sqrt{v_F^2 k^2 + J^2 + \frac{g^2}{4}
    \pm v_F\sqrt{\mathcal{A}(k)}},
\end{align}

where, $\mathcal{A}(k)=\,4J^2 k^2 + (\mathbf{k}\cdot\mathbf{g})^2$ with $\vec{g} = (g_x, g_y, 0)$ and $\rho = \pm 1$, is an index obtained by the diagonalization of the $\vec{\tau}$ subspace obtained after squaring. While it is not apparent, one can show that this dispersion corresponds to the existence of two pairs of Weyl points, a pair at $\vec{k} = (\frac{g_x}{gv_F} \sqrt{J^2 + (g/2)^2}, \frac{g_y}{gv_F} \sqrt{J^2 + (g/2)^2}, 0)$ and a pair at $\vec{k} = 0$ but separated in energy.

\begin{figure}[h]
\centering
\includegraphics[width=0.4\textwidth]{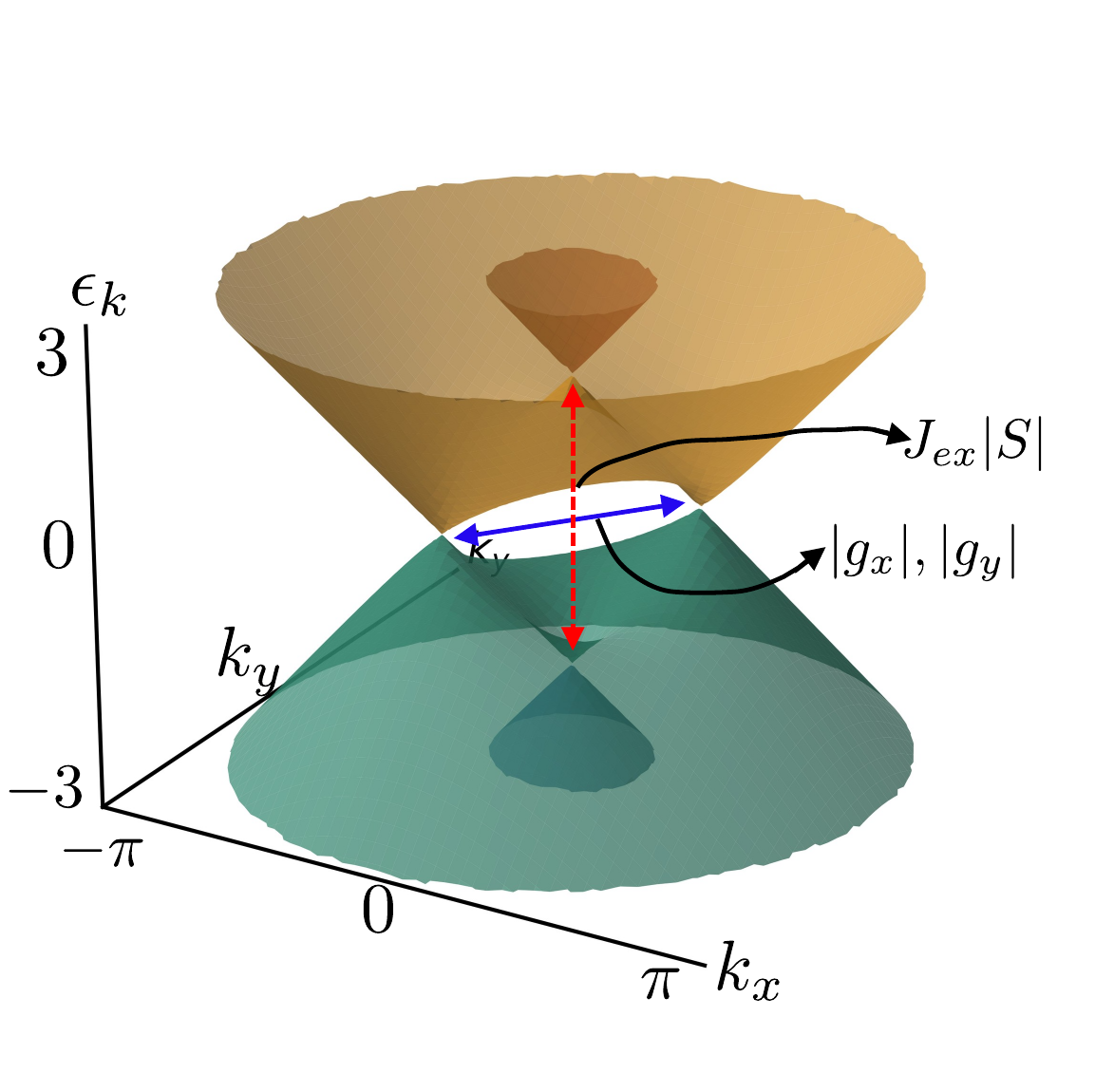}
\caption{Full energy spectrum from Eq.~(\ref{Eq8}) for an inversion-asymmetric Weyl spectrum in the $k_x$–$k_y$ plane with $J_{ex}=1$, $g_x=g_y=\pi/2$, $k_z=0$, and $v_F=1$.}
\label{Fig2}
\end{figure}

We show the spectrum in Fig. \ref{Fig2} where we can see that the four-band system remains untilted. We have verified that the monopole charges of the Weyl nodes in each case numerically, and they work out to be $\pm 1$.

The existence of the two Weyl points separated in momentum space signals the existence of an anomalous Hall current. Naively, the Hall conductivity is determined by the integral of the Berry curvature $\Omega_{ij}$ as 

\begin{align}
\sigma_{ij}
=
-\,\frac{e^{2}}{\hbar}
\sum_{n}
\int \frac{d^{d}k}{(2\pi)^{d}}\;
f_{n}(\mathbf{k})\;
\Omega^{(n)}_{ij}(\mathbf{k}),
\end{align}

where $n$ is a sum over bands, $d$ is the spacetime dimension, and $f(\vec{k})$ is the Fermi-Dirac distribution. It then follows that the Hall conductivity at $T=0$ is given purely by the topological term since there is no tilt to be coupled to the chemical potential. The final expression for the conductivity tensor follows trivially from previous works \cite{TopoS1,Menon2021_II,Menon2021} as 

\begin{align}
    \sigma_{yz} &= \frac{e^2}{2\pi^2\hbar} \frac{g_x}{gv_F} \sqrt{J^2 + (g/2)^2} \nonumber \\
    \sigma_{xz} &= \frac{e^2}{2\pi^2\hbar} \frac{g_y}{gv_F} \sqrt{J^2 + (g/2)^2}, \nonumber \\
    \sigma_{xy} &= 0,
\end{align}

where $g = |\vec{g}| = \sqrt{g_x^2 + g_y^2}$. Noticeably, the results depend on both the pitch vector and the strength of the exchange coupling $J$, suggesting that the origin of this effect is the breaking of {\it time reversal symmetry} by the spin texture. One can see that if either $g_x$ or $g_y$ is zero, the respective Hall conductivity component vanishes, showing that the pitch vector is the driving element in the occurrence of this phenomenon, whereas $J$ adds to the strength of the effect but cannot generate it on its own. Further, if we assume $g_x = g \cos \eta$ and $g_y = g \sin \eta$, then 

\begin{align}
    \sigma_{yz} &= \frac{e^2}{2\pi^2\hbar} \frac{1}{v_F} \sqrt{J^2 + (g/2)^2} \cos \eta \nonumber \\
    \sigma_{xz} &= \frac{e^2}{2\pi^2\hbar} \frac{1}{v_F} \sqrt{J^2 + (g/2)^2} \sin \eta.
\end{align}

We see that there is in a sense a rotational invariance as we change the angle $\eta$ and hence the components of $g$ - the sum in quadrature of the non-zero Hall conductivities is unchanged as $\sqrt{\sigma_{xz}^2 + \sigma_{yz}^2} = \frac{e^2}{2\pi^2\hbar} \frac{1}{v_F} \sqrt{J^2 + (g/2)^2}$. There are, then, circles of constant Hall conductivity in the $(g_x, g_y)$ plane.

In the geometry considered, one can argue that such a bulk analysis might not be valid. To this end, we have explicitly computed the AHE in a lattice model slab of dimension $300 \times 300 \times 20$ and we find that the Hall conductivity takes the form $\sigma_{xy} \propto Q$ (where $2Q$ is the node separation; see supplementary material). In fact, around $80\%$ of the bulk AHE is captured in this setup  up to a few percent in variation depending on the value of the node separation. 

We introduce now a magnetic field in the $z$-direction to study the possibility of the chiral magnetic effect \cite{Zyuzin2012, Menon2021, Chang2015, Chang2015_II, Fukushima2008, Oka2016} in this model. Using the symmetric gauge $\mathbf{A}=(-By/2,\,Bx/2,\,0)$ and defining
$\pi_x = -i\hbar \partial_x + eA_x$,
$\pi_y = -i\hbar \partial_y + eA_y$, we construct the ladder operators
\begin{equation}
    a = \frac{\ell_B}{\sqrt{2}\hbar}(\pi_x - i \pi_y), \qquad
    a^\dagger = \frac{\ell_B}{\sqrt{2}\hbar}(\pi_x + i \pi_y),
\end{equation}
where $\ell_B = \sqrt{\frac{\hbar}{eB}}$ is the magnetic length. The Landau-level spinor ansatz is chosen as
\begin{equation}
    \Psi_{n} =
    \begin{pmatrix}
        u_\uparrow |n-1\rangle\\
        v_\uparrow |n\rangle\\
        u_\downarrow |n-1\rangle\\
        v_\downarrow |n\rangle
    \end{pmatrix},
\end{equation}

\vspace{0.8cm}

ensuring that all appearances of $a$ and $a^\dagger$ reduce to $\sqrt{n}$. Projecting to the Landau level basis with $\Omega_n = v_F \sqrt{2n}\hbar/\ell_B$, we find the that the Hamiltonian may be diagonalized by the squaring trick to yield the eigenvalues. Thus, the energy spectrum for $n \ge 1$ is

\begin{widetext}
\begin{equation}
\epsilon_{n}^{(s,\delta)}(k_z)=
s \sqrt{
v_F^2 k_z^2 + \Omega_n^2 + \frac{g^2}{4} + J_{ex}^2
+ \delta
\sqrt{4 v_F^2 k_z^2 J_{ex}^2 + 4 \Omega_n^2 J_{ex}^2 + g^2 \Omega_n^2}},
\end{equation}
\end{widetext}

 with $s=\pm$ and  $\lambda=\pm$. For $n=0$, 
\begin{equation}
\epsilon_{0}^{s}(k_z) = s|v_F k_z - J_{ex}|.
\end{equation}

\begin{figure}[h]
\centering
\includegraphics[width=0.4\textwidth]{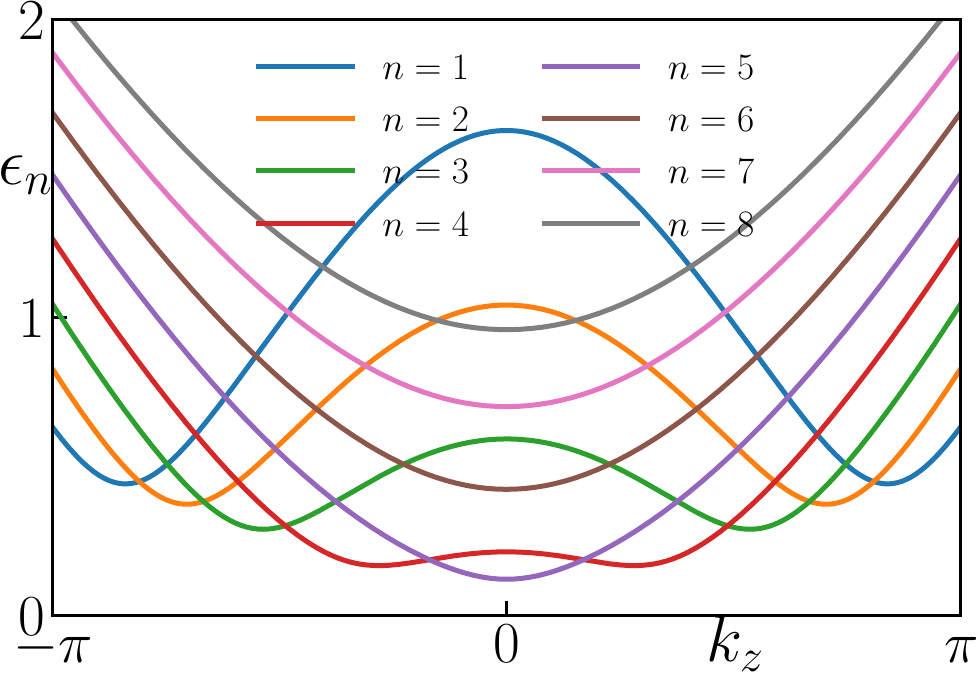}
\caption{
Landau-quantized band structure as a function of the longitudinal momentum $k_z$ for several Landau indices $n$. Each Landau level splits into four branches due to exchange hybridization, labeled by $(s,\delta)$. While higher-index sub-bands exhibit a conventional single-minimum dispersion with a band edge at $k_z=0$, the lower hybridized branch develops a double-minimum structure with degenerate band minima at finite $\pm k_z^\ast$. Parameters used are $J_{ex}=3$, $g=1$, $s=+1$, $\delta=-1$, $l_B=1$ and $\hbar=v_F=1$.}
\label{Fig3}
\end{figure}
 One cannot obtain this result by simply substituting $n=0$ in Eqn.(15) - care needs to be taken to project onto the ground state manifold. 

In a magnetic field, the spectrum (as shown in Fig.\ref{Fig3}) reorganizes into Landau sub-bands that disperse along the field direction $k_z$. Exchange coupling hybridizes the two spin sectors of each Landau level, yielding four branches per index $n$ with energies $\epsilon_n^{(s,\delta)}(k_z)$. While high-index sub-bands exhibit the conventional single-minimum Dirac-like dispersion with a band edge at $k_z=0$, the lower hybridized branch undergoes a qualitative reconstruction at small $n$. Specifically, the competition between the Landau quantization scale $\Omega_n$ and the exchange energy $J_{ex}$ drives a transition from a single minimum at $k_z=0$ to a double-minimum structure with degenerate band minima at finite $\pm k_z^\ast$.

\begin{figure}[h]
\centering
\includegraphics[width=0.4\textwidth]{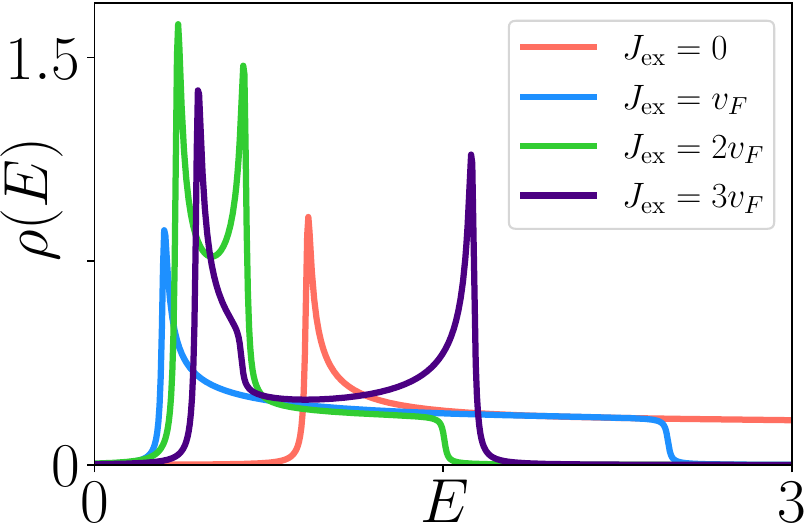}
\caption{Density of states $\rho_n(E)$ for the Landau-quantized spectrum, using the same parameters as in Fig. \ref{Fig3}. The exchange-driven Lifshitz-like reconstruction of the lower hybridized Landau sub-band generates additional one-dimensional van Hove singularities, visible as enhanced features in the DOS.}
\label{Fig3b}
\end{figure}

This exchange-driven reconstruction constitutes a Lifshitz-like transition within an individual Landau sub-band, changing the number of Fermi points at fixed chemical potential without breaking any symmetry. As a consequence, additional one-dimensional van Hove singularities appear in the density of states, which may lead to field-tunable anomalies in quantum oscillations and thermodynamic response. We plot the subband DOS, $\rho_n$ in Fig. \ref{Fig3b}, respectively, to demonstrate this explicitly. 
 
We had introduced the magnetic field in the hope of detecting the CME and to this end, the current response to the applied magnetic field can be calculated via \cite{Menon2021} as

\begin{align}
    j_z = -\frac{e}{2\pi l_B^2} \lim_{\Lambda \rightarrow \infty} \int_{-\Lambda}^{\Lambda} \frac{dk_z}{2\pi} \frac{\partial}{\partial z} \Bigg[ \epsilon_{0}^{-} + \sum_{n=1}^{\infty} \sum_{s=\pm} \epsilon_{n}^{s,-} \Bigg].
\end{align}

The contribution from the higher Landau levels vanish by symmetry and we are simply left with the contribution of the lowest LL. The derivative and the integral annihilate leaving us to evaluate the limit appropriately. 

\begin{align}
    j_z = -\frac{e^2B}{4\pi^2} [\epsilon_0^{-}(\Lambda) - \epsilon_0^{-}(-\Lambda)]
\end{align}

Inserting the appropriate energy terms into the expression above, we find that $\epsilon_0^{-}(\Lambda) - \epsilon_0^{-}(-\Lambda) = -|v_F\Lambda - J| +  |-v_F \Lambda - J| = -v_F\Lambda + J + v_F\Lambda + J = 2J$. Thus, the chiral magnetic parameter $\mu_5^{\rm{eff}}$ is given by

\begin{align}
    \mu_5^{\rm{eff}} = \frac{j_z}{B} = -\frac{e^2}{2\pi^2}J_{ex}.
\end{align}

We emphasize that the CME-like response discussed here refers to the effective low-energy response associated with the induced Weyl structure, not a generic claim of equilibrium chiral magnetic current in a lattice system. The CME is manifestly a 3d effect, and hence, one might also question its validity in the present setup, much like the AHE. In this case as well, we have simulated the CME on a lattice geometry of $300\times300\times20$ and we find that $\mu_5^{\rm{eff}} \propto \Delta E$, where $\Delta E$ is the energy separation of the nodes. In fact, we also find that the CME in this geometry is enhanced by a few percent as compared to the bulk $300\times300\times300$ geometry - see supplementary materials.

\noindent {\it Time Dependent Pitch Vector}. We now consider the case where the pitch vector has time-dependence, and it takes the form $\phi(r,t) = g[x \cos \omega t + \lambda y\sin \omega t]$. The algebra for the unitary transformation changes accordingly and is $H \rightarrow H^{'} = U^{\dagger}HU -iU^{\dagger}\partial_t U = U^{\dagger}HU - \frac{1}{2}(\partial_t \phi) \sigma_z$. Thus, the transformed Hamiltonian takes the form

\begin{align}
    H^{'}(r,t) &= v_F\Bigg[ \tau_x \Big(k_x - \frac{1}{2}g\cos \omega t \sigma_z \Big) \nonumber \\ &+ \tau_y \Big(k_y - \frac{\lambda}{2}g\sin \omega t \sigma_z \Big)+\tau_zk_z \Bigg] \nonumber \\
    &~ + J_{ex}|\vec{S}|\sigma_x + \frac{\omega g}{2}\Big[x\sin \omega t - \lambda y \cos \omega t \Big]\sigma_z.
\end{align}

While it may look like this Hamiltonian is equivalent to a DSM under the influence of circularly polarized light and a Zeeman field, we remind the reader that $\vec{\nabla} \phi$ is curl-less. Floquet systems have been studied previously in the context of topological semimetals \cite{Floquet_WSM,Menon_2018,Menon_2020}, although the results we present here are novel. We may now choose a regime where $\omega \gg \Delta E$, where $\Delta E$ is the bandwidth of the system and then invoke the van-Vleck expansion given by

\begin{align}
    H_{\rm{eff}} = H_0^F + \frac{[H^F_{-1},H^F_{+1}]}{\omega} + \mathcal{O}(\frac{1}{\omega^2}),
\end{align}

to leading order in perturbation theory, where $H_n^F = \frac{1}{T}\int_0^T H(t) e^{in\omega t} dt$. It may appear that the $U\partial_t U$ term may lead to position dependencies in the effective Hamiltonian, but a simple calculation shows that this dependence is removed. The effective Floquet Hamiltonian in this setup is then given by 

\begin{align}
    H_{\rm{eff}} &= v_F\Big[ \tau_xk_x + \tau_yk_y + \tau_z(k_z - \lambda \frac{v_F^2g^2}{4\omega}) \Big] \nonumber \\ &+ J_{ex}|\vec{S}|\sigma_x.
\end{align}

At leading order in the high-frequency (van Vleck) expansion, the effective Hamiltonian is translation invariant and may be diagonalized straightforwardly,

\begin{align}
\epsilon_{s,t}(k) = sJ_{ex}|\vec{S}| + tv_F\sqrt{k_x^2 +k_y^2 + (k_z - \lambda \frac{v_F^2g^2}{4\omega})^2},
\end{align}

where $s,t = \pm$ gives the four distinct bands. Then we can see that when $s \cdot t <0$, there is the existence of a nodal sphere at $E = 0$ specified by the equation $k_x^2 +k_y^2 + (k_z - \lambda \frac{v_F^2g^2}{4\omega})^2 = J_{ex}^2\vec{S}^2$. 

\begin{figure}[h]
\centering
\includegraphics[width=0.4\textwidth]{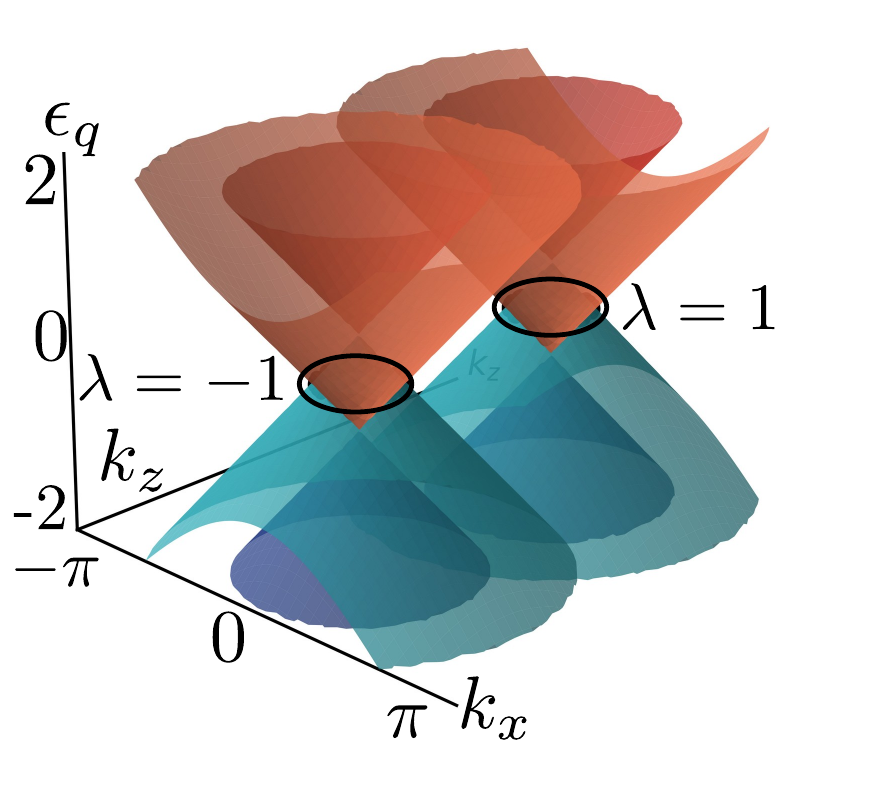}
\caption{The Floquet bulk spectrum of a nodal-ring semimetal for $\lambda=\pm 1$ in the $k_x$–$k_z$ plane is shown, resulting from coupled temporal and spatial spin-texture dynamics in a three-dimensional orbital Dirac semimetal. Parameters used are $J_{ex}=0.5$, $g_x=g_y=2.5$, $k_y=0$, $\omega=1$, and $v_F=1$.  }
\label{Fig4}
\end{figure}

We now analyze the structure of the van Vleck expansion beyond leading order. While higher-order terms can generate additional operator contributions, the algebraic structure of the Fourier components imposes strong constraints on the form of the effective Hamiltonian. In particular, the grading structure under conjugation by $\sigma_x$ restricts the allowed operator content and forbids conventional mass terms within this subspace. As a result, the leading-order degeneracy is not lifted through standard gap-opening mechanisms.

From the explicit structure of Eq.~(20), we note that the static component $H_0$ contains the Dirac Hamiltonian together with the exchange term proportional to $\sigma_x$, while all nonzero Fourier components arise from the time-dependent pitch vector and are proportional to $\sigma_z$.

To make this structure precise, it is useful to introduce a $\mathbb{Z}_2$ grading of operators under conjugation by $\sigma_x$:
\begin{equation}
\sigma_x A \sigma_x = \pm A, \qquad A \in \mathcal{A}_\pm,
\end{equation}
where $\mathcal{A}_+$ ($\mathcal{A}_-$) denotes the set of operators that are even (odd) under this transformation. One readily verifies that
\begin{equation}
H_0 \in \mathcal{A}_+, \qquad H_n \in \mathcal{A}_- \quad (n \neq 0).
\end{equation}

The van Vleck effective Hamiltonian is expressed as a series of nested commutators of the Fourier components, subject to the constraint that the total harmonic index vanishes. Up to second order, for instance, one has
\begin{align}
H_{\mathrm{eff}} = H_0 + \frac{1}{\omega} \sum_{n \neq 0} \frac{[H_{-n}, H_n]}{n}
\nonumber \\ + \frac{1}{\omega^2} \sum_{n,m \neq 0} \frac{[H_{-n}, [H_{n-m}, H_m]]}{nm} + \cdots.
\end{align}
All higher-order terms share this same structure.

The crucial point is that the above $\mathbb{Z}_2$ grading is preserved under commutation:
\begin{equation}
[\mathcal{A}_+, \mathcal{A}_+] \subset \mathcal{A}_+, \quad
[\mathcal{A}_+, \mathcal{A}_-] \subset \mathcal{A}_-, \quad
[\mathcal{A}_-, \mathcal{A}_-] \subset \mathcal{A}_+.
\end{equation}
Since each nonzero harmonic $H_n$ lies in $\mathcal{A}_-$, any contribution to the static effective Hamiltonian must involve an even number of such terms in order to satisfy the zero-harmonic constraint. It follows that every term generated in the van Vleck expansion belongs to $\mathcal{A}_+$, and therefore
\begin{equation}
H_{\mathrm{eff}} \in \mathcal{A}_+ \quad \text{to all orders in } \omega^{-1}.
\end{equation}

This restriction has an immediate consequence: the effective Hamiltonian can only contain operators that are even under conjugation by $\sigma_x$, with no terms proportional to $\sigma_y$, $\sigma_z$, or $\tau_i \sigma_{y,z}$. As a result, $H_{\mathrm{eff}}$ is block-diagonal in the eigenbasis of $\sigma_x$, and no mass term is generated.

\begin{figure}[h]
\centering
\includegraphics[width=0.4\textwidth]{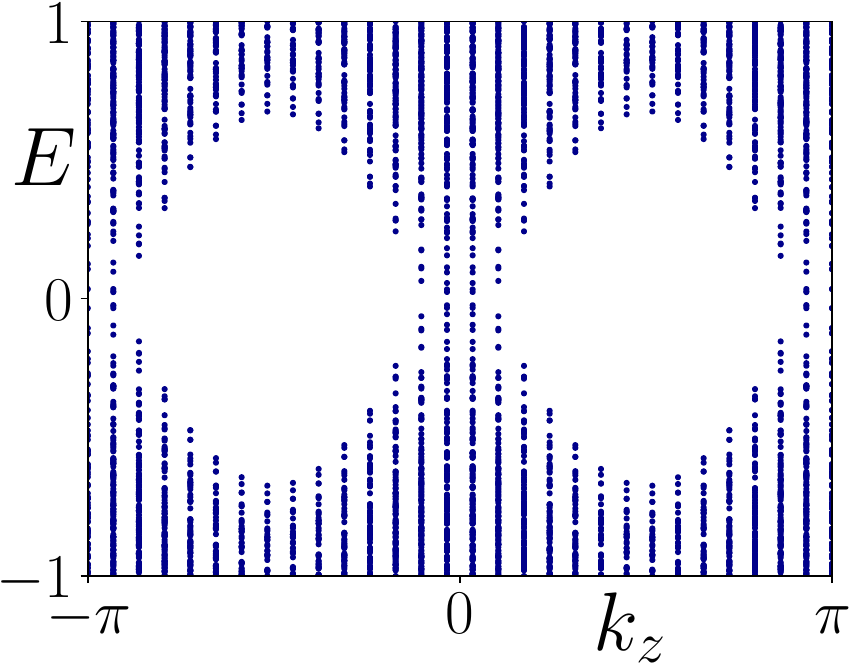}
\caption{Quasienergy spectrum from the truncated Floquet (Sambe) Hamiltonian (one-photon sector) for the lattice-regularized model, plotted as a function of $k_z$. A closed quasienergy degeneracy structure is observed near zero energy, forming a descendant of the nodal sphere obtained at leading order in the van Vleck expansion. Parameters: $v_F=1, \lambda = 1 ,g = 0.5, J_ex=0.5v_F, \omega=1000$.}
\label{Fig5}
\end{figure}

To corroborate the van Vleck analysis beyond perturbation theory, we compute the Floquet spectrum directly in a truncated Sambe space. Restricting to the one-photon sector, we construct the block Floquet Hamiltonian including the $m=0,\pm1$ harmonics and diagonalize it numerically. The resulting quasi-energy spectrum, shown in Fig. \ref{Fig5}, exhibits a closed degeneracy structure in quasi-energy space that is continuously connected to the nodal sphere obtained at leading order. We note that, due to the presence of coordinate-dependent terms in the Floquet harmonics, translational symmetry in the transverse directions is explicitly broken, and $k_x, k_y$ are not good quantum numbers in this geometry. This suggests that, in the absence of transverse translational symmetry, the nodal sphere reorganizes into a quantized degeneracy shell in transverse-mode space.

{\it Conclusions.} In this work, we examined the effects of coupling a spin-texture to an orbital Dirac semimetal. We employed a unitary transformation to find a Hamiltonian whose spectrum and eigenstates were controlled by the pitch vector and exchange coupling. Using a linear pitch vector led to the formation of four Weyl nodes, two at distinct momenta space and the other two split in energy. This novel system exhibited the classic anomalous Hall and chiral magnetic effects characteristic of Weyl systems. The topological Hall response exhibited a circle of invariance in the $(g_x,g_y)$ plane, whereas the CME remained proportional to the exchange coupling $J_{ex}$. Along the way, we found the magnetic-field-induced Landau level structure exhibited a Lifshitz-like transition as we increased the Landau index $n$. We then chose a time-dependent pitch vector and analyzed the resulting system using van-Vleck perturbation theory. We found that the leading-order Floquet effective Hamiltonian hosts a nodal sphere in momentum space. Beyond this approximation, the driven system exhibits a closed quasi-energy degeneracy structure, constrained by the operator algebra of the Floquet expansion and continuously connected to the nodal sphere. The novel results of this work invite experimental exploration further supported by the recent discovery of the material family Pr$_8$CoGa$_3$ which hosts orbital Dirac fermions. 

AM and PC contributed equally to this project. Pritam Chatterjee acknowledges Prof. Arijit Saha, Prof. Kenji Fukushima, and Prof. Takashi Oka for useful discussions. AM would like to thank NCTS Taiwan for their continuing support of his postdoctoral work.

\bibliography{bibfile}{}

\end{document}